\documentclass[12pt]{article}
\usepackage{amsfonts,bm}
\usepackage{graphicx}
\baselineskip
\offinterlineskip
\newcommand{\n}{\newline}

\begin{document}

\begin{center}
{\bf A Sequence of Quantum Gates}
\end{center}

\begin{center}
{\bf Yorick Hardy$^\ast$ and Willi-Hans Steeb$^\dag$} \\[2ex]

$\ast$
Department of Mathematical Sciences, \\
University of South Africa, Pretoria, South Africa, \\
e-mail: {\tt hardyy@unisa.ac.za}\\[2ex]

$\dag$
International School for Scientific Computing, \\
University of Johannesburg, Auckland Park 2006, South Africa, \\
e-mail: {\tt steebwilli@gmail.com}\\[2ex]
\end{center}

\strut\hfill

{\bf Abstract} We study a sequence of quantum gates
in finite-dimensional Hilbert spaces given by
the normalized eigenvectors of the unitary operators. The corresponding 
sequence of the Hamilton operators is also given. From the Hamilton 
operators we construct another hierarchy of quantum gates 
via the Cayley transform.

\strut\hfill

\section{Introduction}

In the finite dimensional Hilbert space ${\mathbb C}^n$ 
quantum gates are given by $n \times n$ unitary matrices \cite{1,2,3,4,5}.
For example for $n=2$ we have the phase gate, NOT-gate
and Hadamard gate  
$$
U_P = \pmatrix { 1 & 0 \cr 0 & e^{i\phi} }, \quad
U_{NOT} = \pmatrix { 0 & 1 \cr 1 & 0 }, \quad
U_H = \frac1{\sqrt2} \pmatrix { 1 & 1 \cr 1 & -1 } \,.
$$
For the Mach-Zehnder interferometer these quantum gates play 
the central role \cite{5}. The Pauli spin matrices 
$\sigma_x$, $\sigma_y$, $\sigma_z$ with $\sigma_x=U_{NOT}$ are 
also quantum gates. In higher dimensions we have the CNOT gate,
Toffoli gate, Fredkin gate
$$
U_{CNOT} = I_2 \oplus \sigma_x, \quad
U_T = I_6 \oplus \sigma_x, \quad 
U_F = (1) \oplus \sigma_x \oplus I_5 
$$
where $\oplus$ denotes the direct sum.
Teleportation in the Hilbert space ${\mathbb C}^8$ is provided
by the unitary matrix \cite{5}
$$
U = U_8 U_7 U_6 U_5 U_4 U_3 U_2 U_1
$$
where 
\begin{eqnarray*}
U_1 = I_2 \otimes U_H \otimes I_2, & \quad &
U_2 = I_2 \otimes U_{CNOT}, \\
U_3 = U_{CNOT} \otimes I_2, & \quad &
U_4 = U_H \otimes I_2 \otimes I_2, \\
U_5 = I_2 \otimes U_{CNOT}, & \quad &
U_6 = I_2 \otimes I_2 \otimes U_H, \\
U_7 = I_4 \oplus U_{NOT} \oplus U_{NOT}, & \quad &
U_8 = I_2 \otimes I_2 \otimes U_H \,.
\end{eqnarray*}
The $n \times n$ unitary matrices form the compact Lie group $U(n)$.
If $\det(U)=1$ we have the Lie subgroup $SU(n)$. 
The eigenvalues of a unitary matrix are of the form 
$e^{i\phi_j}$ $(j=1,2,\dots,n)$, where $\phi_j \in [0,2\pi)$, i.e.
they are elements of the commutative Lie group $U(1)$. 
\newline

Beside the Hilbert space ${\mathbb C}^n$ we also utilize the Hilbert space 
of $n \times n$ matrices over $\mathbb C$ with the scalar product
$$
\langle A,B \rangle := \mbox{tr}(AB^*) \,.
$$
This implies a norm $\|A\|^2=\mbox{tr}(AA^*)$ and a distance measure 
(Hilbert Schmidt distance)
$$
\|A-B\| = \mbox{tr}((A-B)(A-B)^*) = 
\mbox{tr}(AA^*) - \mbox{tr}(AB^*) - \mbox{tr}(BA^*) + \mbox{tr}(BB^*) \,.
$$
In particular if we have two $n \times n$ unitary matrices, say
$U$ and $V$ then 
$$
\|U-V\| = 2n - \mbox{tr}(UV^*) - \mbox{tr}(VU^*) 
$$
since $\mbox{tr}(UU^*)=n$, $\mbox{tr}(VV^*)=n$. Consider the 
two $n \times n$ unitary matrices $U$ and $e^{i\phi}U$. Then
$$
\|U-e^{i\phi}U\| = 2n - 2n\cos\phi
$$
where $\phi \in [0,2\pi)$. The norm takes a minimum for
$\phi=0$. We consider two unitary matrices $U$ and $V$ equivalent 
if and only if $V=e^{i\phi}U$ for $\phi \in [0,2\pi)$.
This leads to equivalence classes. Thus for example
$\sigma_y$ and $i\sigma_y$ belong to the same equivalence class. 
We define the distance measure \cite{6}
$$
d(U,V) := \min_{\phi \in [0,2\pi)} 
\frac1{\sqrt{2n}} \|e^{i\phi}U-V\| \,.
$$
We ask the following question. Given a unitary matrix $U_0$
(not of the form $e^{i\phi}I_n$, where $I_n$ is the identity matrix).
We calculate the eigenvalues and corresponding normalized eigenvectors
of $U_0$. Using the normalized eigenvectors as columns we form a new unitary
matrix $U_1$. This process is repeated and thus provides a sequence
of unitary matrices. Now every unitary matrix $U$ can be written as 
$U=\exp(K)$, where $K$ is a skew-hermitian matrix. $K$ can be identified
with a Hamilton operator (hermitian matrix $H$) via 
$K=-iH$. Thus we find the corresponding sequence of Hamilton operators
$H_0$, $H_1$, $\dots$. 
\newline

From these hermitian matrices we can also form a new hierarchy of unitary matrices
via the Cayley transform. The Cayley transform for an $n \times n$ hermitian 
matrix $H$ is given by
$$
V = (H-iI_n)(H+iI_n)^{-1} 
$$
where $V$ is a unitary matrix. Note that $+1$ cannot be an eigenvalue of 
the unitary matrix $V$.     
\newline

\section{Sequence of Unitary Matrices} 

Let $\{\,\mathbf{e}_{1,n},\,\ldots,\,\mathbf{e}_{n,n}\,\}$ denote the
standard basis in $\mathbb{C}^n$. Let $\lhd$ denote an ordering
on $\mathbb{C}^n$ and let $F_n:U(n)\to U(n)$ be a function with
the properties that $F_n(U)\mathbf{e}_{j,n}$ is an eigenvector of
$U$ for $j=1,2,\ldots,n$ and the columns of $F_n(U)$ are in order
with respect to $\lhd$, i.e.
$$
F_n(U)\mathbf{e}_{1,n} \lhd
F_n(U)\mathbf{e}_{2,n} \lhd \cdots \lhd
F_n(U)\mathbf{e}_{j,n}.
$$
Consider a sequence $U_0$ (given), $U_1$, $U_2$, \ldots of
unitary matrices acting on $\mathbb{C}^n$ defined by
$$ 
U_{k+1}:=F_n(U_k). 
$$
We have the spectral decompositions
$$ 
U_k := F_n(U_k)\,
\textrm{diag}\left(
e^{i\theta_{k,1}}, e^{i\theta_{k,2}},\ldots,e^{i\theta_{k,n}}
\right)F_n(U_k)^* 
$$
where $e^{i\theta_{k,j}}$ is the $j$-th eigenvalue of $U_k$ with
corresponding eigenvector $F_n(U_k)\mathbf{e}_{j,n}$.
It follows that the corresponding Hamilton operators are
$$ 
H_k := F_n(U_k)\,
\textrm{diag}\left(-\theta_{k,1}, -\theta_{k,2},\ldots,-\theta_{k,n}
\right)F_n(U_k)^* 
$$
and the Cayley transforms are given by
$$
V_k := -F_n(U_k)\,
\textrm{diag}\left(
\frac{i+\theta_{k,1}}{i-\theta_{k,1}},
\frac{i+\theta_{k,2}}{i-\theta_{k,2}}, \ldots,
\frac{i+\theta_{k,n}}{i-\theta_{k,n}}\right)F_n(U_k)^*.
$$
Suppose that the sequence converges in the sense of the
Hilbert Schmidt distance. Then
$$
\lim_{k\to \infty} U_{k+1}^*U_k = 
\lim_{k\to\infty}
\textrm{diag}\left(
e^{i\theta_{k,1}},e^{i\theta_{k,2}},\ldots,e^{i\theta_{k,n}}
\right)F_n(U_k)^* = I_n
$$
from which follows
$$
\lim_{k\to\infty} U_k = \lim_{k\to\infty} F_n(U_k) =
\lim_{k\to\infty} \textrm{diag}\left(
e^{i\theta_{k,1}},e^{i\theta_{k,2}},\ldots,e^{i\theta_{k,n}}\right).
$$
Thus for convergence we must have
$$
e^{i\alpha_1}\mathbf{e}_{1,n}
\lhd e^{i\alpha_2}\mathbf{e}_{2,n}\lhd\cdots\lhd e^{i\alpha_n}
\mathbf{e}_{n,n}
$$
for all $\alpha_1,\alpha_2,\ldots,\alpha_n\in\mathbb{R}$.\\

We note that even though the eigenvalues need never explicitly
appear in the definition of $F_n$, they still play a role in the
convergence of the sequence.

\section{Examples}

To define the sequence we need
\begin{enumerate}
 \item the ordering $\lhd$,
 \item the function $F_n$,
 \item and the initial unitary matrix $U_0$.
\end{enumerate}

We consider a few different $U_0$. For each of
the sequences we will use the same ordering and the same definition
of $F_n$ (dependent on $n$).\\

{\bf Example $\lhd$.}\\

The ordering $\lhd$ we consider is 
$$
\mathbf{x}\lhd\mathbf{y}
\Leftrightarrow
\left(|\mathbf{e}_{u(\mathbf{x},\mathbf{y})}^*\mathbf{x}|
>|\mathbf{e}_{u(\mathbf{x},\mathbf{y})}^*\mathbf{y}|\right)
\vee
\left(|\mathbf{e}_{u(\mathbf{x},\mathbf{y})}^*\mathbf{x}|
       =|\mathbf{e}_{u(\mathbf{x},\mathbf{y})}^*\mathbf{y}|
        \wedge
        \arg(\mathbf{e}_{u(\mathbf{x},\mathbf{y})}^*\mathbf{x})
         <\arg(\mathbf{e}_{u(\mathbf{x},\mathbf{y})}^*\mathbf{y})\right)$$
where
$$
u(\mathbf{x},\mathbf{y}):=\min\left\{v\in\{1,\ldots,n\}\,:\,
\mathbf{e}_v^*\mathbf{x}\neq\mathbf{e}_v^*\mathbf{y}\right\}.
$$

{\bf Example $F_n$.}\\

For each of the sequences we will use the same definition
of $F_n$ (dependent on $n$). We choose $F_n(U)$ as follows.
Let $\{\mathbf{x}_1,\mathbf{x}_2,\ldots,\mathbf{x}_n\}$ be
an orthonormal set of eigenvectors of $U$. The eigenvectors
are determined as follows. Let $\Pi_\lambda$ denote the projection operator
onto an eigenspace of $U$ for the eigenvalue $\lambda$.
Then the Gram-Schmidt orthonormalization process applied to
$\{\,\Pi_\lambda{\bf e}_{1,n},\,\Pi_\lambda{\bf e}_{2,n},\,\ldots
      \Pi_\lambda{\bf e}_{n,n}\,\},$
in that order (and discarding zero vectors), yields an orthonormal
basis $\{\mathbf{x}_j,\mathbf{x}_{j+1},\ldots\}$
for that eigenspace.
Define
$$
\mathbf{y}_{k}:=\frac{|\mathbf{e}_{l(\mathbf{x}_k),n}^*\mathbf{x}_{k}|}
    {\mathbf{e}_{l(\mathbf{x}_k),n}^*\mathbf{x}_{k}}\mathbf{x}_{k},\qquad
l(\mathbf{x}):=\min\left\{ v\in\{1,\ldots,n\}\,:\,
  \mathbf{e}_{v,n}^*\mathbf{x}_{k}\neq 0\right\}.
$$
Now let $\sigma:\{1,2,\ldots,n\}\to\{1,2,\ldots,n\}$ be the permutation
such that
$$
\mathbf{y}_{\sigma(1)}\lhd
\mathbf{y}_{\sigma(2)}\lhd
\cdots\lhd\mathbf{y}_{\sigma(n)}.
$$
Then we define
$$ 
F_n(U):=\sum_{j=1}^n {\bf y}_{\sigma(j)}\mathbf{e}_{j,n}^*.
$$
{\bf Example 1.1.}\\

As a first example we consider the $U_0=U_{NOT}=\sigma_x$ gate. 
This is a special case of example 1.2 below.
The corresponding Hamilton operator $H_0$ is 
$$
H_0 = -\frac{\pi}2 \pmatrix { 1 & -1 \cr -1 & 1 }  
$$
and the unitary operator given by the Cayley transform of $H_0$ is
$$
V_0 = -\frac{1}{1+i\pi}\pmatrix { 1 & i\pi \cr i\pi & 1 } 
$$
with the eigenvalues $+i$ and $-i$. The eigenvalues of $U_0$ are $+1$ and $-1$ 
and the corresponding normalized eigenvectors are
$$
\frac1{\sqrt2} \pmatrix { 1 \cr 1 }, \qquad
\frac1{\sqrt2} \pmatrix { 1 \cr -1 } \,.
$$
This leads us to the Hadamard gate
$$
U_1 = \frac1{\sqrt2} \pmatrix { 1 & 1 \cr 1 & -1 } \,.
$$
The eigenvalues of $U_1$ are $+1$ and $-1$ with the corresponding
normalized eigenvectors 
$$
\frac1{\sqrt8} \pmatrix { \sqrt{4+2\sqrt2} \cr \sqrt{4-2\sqrt2} }, \qquad
\frac1{\sqrt8} \pmatrix { \sqrt{4-2\sqrt2} \cr -\sqrt{4+2\sqrt2} } \,.
$$
The corresponding Hamilton operator is
$$
H_1 = -\frac{\pi}8 \pmatrix { 4-2\sqrt2 & -\sqrt8 \cr -\sqrt8 & 4+2\sqrt{2} }
$$
with the Cayley transform of $H_1$
$$
V_1 = \frac{1}{2(\pi+i)}\pmatrix { -\sqrt2\pi - 2i & -\sqrt2\pi \cr -\sqrt2{\pi} & \sqrt2\pi - 2i } \,.
$$
At the next step we find
$$
U_2 = \frac{1}{\sqrt8}
\pmatrix { \sqrt{4+2\sqrt2}&\sqrt{4-2\sqrt2}\cr
           \sqrt{4-2\sqrt2}&-\sqrt{4+2\sqrt2}}
$$
and ($\alpha:=2+\sqrt2$, $\beta:=2-\sqrt2$)
$$
H_2 = \frac{-\pi\beta}{8+4\sqrt{\alpha}} 
\pmatrix { 1 & (2+\sqrt{\alpha})/\sqrt{\beta} \cr 
(2 + \sqrt{\alpha})/\sqrt{\beta} & 
(4\sqrt{\alpha} + \sqrt2 + 6)/\beta } \,.
$$
The distances between the unitary matrices are
$$
\|U_0-U_1\| = \sqrt{4-2\sqrt2}, \qquad 
\|U_1-U_2\| = \sqrt{4-\sqrt{4+2\sqrt2}-\sqrt{4-2\sqrt2}} \,.
$$
From example 1.2 we find 
$$
\lim_{k\to\infty} U_k=\pmatrix{1&0\cr0&-1}.
$$

{\bf Example 1.2.}\\

Consider the unitary matrix
$$ 
U_0 = \pmatrix{a&b\cr b&-a}
$$
where $a,b\in\mathbb{R}$, $b\neq0$ and $a^2+b^2=1$.
If $a>0$ then we find
$$
U_1 = \pmatrix{a' & b' \cr \mathrm{sgn}(b)b' & -\mathrm{sgn}(b)a' },
\qquad a':=\sqrt{\frac{1+a}{2}},\quad b':=\sqrt{\frac{1-a}{2}},
$$
and for $a<0$ we have
$$
U_1 = \pmatrix{a' & b' \cr -\mathrm{sgn}(b)b' & \mathrm{sgn}(b)a' },
\qquad a':=\sqrt{\frac{1-a}{2}},\quad b':=\sqrt{\frac{1+a}{2}},
$$
while for $a=0$ we find that $U_1$ is the Hadamard gate
$$
U_1 = \pmatrix{a' & b' \cr b' & -a' },
\qquad a':=\frac1{\sqrt2},\quad b':=\frac1{\sqrt2}.
$$
In each case, by construction, we have $a'>0$ and $b'\neq 0$.
We also note that for $a\geq0$ we have $a'>|a|$ and $b'<|b|$.
Since $a',b'\in(0,1]$ we find that sequences beginning with
$$ 
U_0 = \pmatrix{a&b\cr b&-a}
$$
where $a\geq 0$, converge to $\sigma_z$.

{\bf Example 2.}\\

As our second example we consider the Pauli matrix 
$$
U_0 = \sigma_2 = \pmatrix { 0 & -i \cr i & 0 }
$$
with 
$$
H_0 = -\frac{\pi}{2} \pmatrix { 1 & i \cr -i & 1 }, \qquad
V_0 = -\frac1{1+i\pi} \pmatrix { 1 & \pi \cr -\pi & 1 } \,.
$$
The eigenvalues of $U_0$ are $1$ and $-1$ and we obtain 
the unitary matrix
$$
U_1 = \frac1{\sqrt2} 
\pmatrix { 1 & 1 \cr -i & i } \,.
$$
The eigenvectors of $U_1$ are
$$
\frac1{\sqrt{12+4\sqrt3}}\pmatrix{2\cr -(1-i)(\sqrt3+1)},\qquad
\frac1{\sqrt{12-4\sqrt3}}\pmatrix{2\cr (1-i)(\sqrt3-1)} \,.
$$
Thus
$$
U_2 = \frac1{4\sqrt6}
\pmatrix{2\sqrt{12-4\sqrt3} & 2\sqrt{12+4\sqrt3} \cr
          (1-i)\sqrt{24+8\sqrt3}  & (1-i)\sqrt{24-8\sqrt3} }.
$$

{\bf Example 3.}\\

As last example we consider the $3 \times 3$ matrix
$$
U_0 = \pmatrix { 1/\sqrt2 & 0 & 1/\sqrt2 \cr 0 & 1 & 0 \cr 
                 1/\sqrt2 & 0 & -1/\sqrt2 } \,. 
$$
Then 
$$
H_0 = -\frac{\pi}8
\pmatrix { 4-2\sqrt2 & 0 &-\sqrt8 \cr 0 & 0 & 0 \cr-\sqrt8 & 0 & 4+2\sqrt{2} },
$$
$$
V_0 = \frac{1}{2(\pi+i)}
\pmatrix{ -\sqrt2\pi - 2i & 0 & -\sqrt2\pi \cr
         0 & -2(\pi+i) & 0 \cr -\sqrt2{\pi} & 0 & \sqrt2\pi - 2i } \,.
$$
The eigenvalues of $U_0$ are $+1$ and $-1$ with the corresponding
normalized eigenvectors 
$$
\frac1{\sqrt8} \pmatrix { \sqrt{4+2\sqrt2} \cr 0 \cr \sqrt{4-2\sqrt2} }, \qquad
\pmatrix { 0 \cr 1 \cr 0 }, \qquad
\frac1{\sqrt8} \pmatrix { \sqrt{4-2\sqrt2} \cr 0 \cr -\sqrt{4+2\sqrt2} } \,.
$$
For the first step in the sequence we find
$$
U_1 = \frac1{\sqrt8}
      \pmatrix { \sqrt{4+2\sqrt2} & \sqrt{4-2\sqrt2} & 0\cr
                 0 & 0 & \sqrt8 \cr
                 \sqrt{4-2\sqrt2} & -\sqrt{4+2\sqrt2} & 0 } \,.
$$

\section{Kronecker Product, Direct Sum and Star Product}

From the $2 \times 2$ case we can easily construct these
sequences for the $4 \times 4$ case given by the direct sum,
Kronecker product and star product. The star product
of two $2 \times 2$ matrices $A$ and $B$ is defined by
$$
A \star B := 
\pmatrix { a_{11} & 0 & 0 & a_{12} \cr 
0 & b_{11} & b_{12} & 0 \cr
0 & b_{21} & b_{22} & 0 \cr
a_{21} & 0 & 0 & a_{22} } \,.
$$
Note that if $A$ and $B$ are unitary, then $A \star B$ is
unitary. We also have $\det(A \star B)=\det(A)\det(B)$
and $\mbox{tr}(A \star B) = \mbox{tr}(A) + \mbox{tr}(B)$.
\newline

In particular if $F_n$ distributes over any of these operations i.e. 
\begin{eqnarray*}
F_{mn}(U_A\otimes U_B) &=& F_m(U_A)\otimes F_n(U_B), \\
F_{m+n}(U_A\oplus U_B) &=& F_m(U_A)\oplus F_n(U_B), \\
F_{m+n}(U_A\star U_B) &=& F_m(U_A)\star F_n(U_B)
\end{eqnarray*}
for $U_A\in U(m)$ and $U_B\in U(n)$,
then the sequence for $U_0:=U_{0,0}\otimes U_{0,1}$ etc.
is completely determined by the sequence for $U_{00}$ and $U_{01}$.
In other words if $F_{mn}(U_A\otimes U_B)=F_m(U_A)\otimes F_n(U_B)$
then we construct the sequences for $U_{0,0}\in U(m)$ and $U_{0,1}\in U(n)$
$$
U_{0,0},\quad U_{1,0}=F_m(U_{0,0}),\quad U_{2,0}=F_m(U_{1,0}),
\quad\ldots
$$
$$
U_{0,1},\quad U_{1,1}=F_n(U_{0,1}),\quad U_{2,1}=F_n(U_{1,1}),
\quad\ldots
$$
and
$$
U_0, \quad
U_1=F_{mn}(U_0)=U_{1,0}\otimes U_{1,1},\quad
U_2=F_{mn}(U_2)=U_{2,0}\otimes U_{2,1},\quad\ldots
$$

\section{Conclusion}

We defined a sequence of unitary operators on the Hilbert space $\mathbb{C}^n$.
The eigenvectors of each unitary operator is used to construct the next. We
found some necessary criteria for convergence.\\

Some open questions remain. The role of the eigenvalues in the convergence
of the sequence has not been determined. Due to the implicit role of the
eigenvalues we have not examined the sequence of eigenvalues and their
convergence properties.
\newline

Obviously this construction can also be applied to other invertible
matrices. For example the matrix $(\alpha \in {\mathbb R})$
$$
A(\alpha) = \pmatrix { \cosh(\alpha) & \sinh(\alpha) \cr 
                       \sinh(\alpha) & \cosh(\alpha) } 
$$ 
is an element of the non-compact Lie group $SO(1,1)$
(Lorentz transformation). The matrix is also hermitian. 
Applying the construction described above
we find the Hadamard gate $U_H$ which is an element of the
compact Lie group $U(2)$ \cite{7}. Since $A(\alpha)$ is hermitian
we can apply the Cayley transform and find the quantum gate
$$
U(\alpha) = \frac1{\cosh(\alpha)} 
\pmatrix { -i & \sinh(\alpha) \cr \sinh(\alpha) & -i } 
$$   
with $U(\alpha \to \infty) = \sigma_x = U_{NOT}$. 

\strut\hfill

\section*{Acknowledgements}

The first author is supported by the National Research Foundation (NRF),
South Africa. This work is based upon research supported by the National
Research Foundation. Any opinion, findings and conclusions or recommendations
expressed in this material are those of the author(s) and therefore the
NRF do not accept any liability in regard thereto.

\strut\hfill

\end{document}